\documentclass[prl,aps,twocolumn,superscriptaddress,
]{revtex4}
\usepackage{amsmath,amssymb,graphicx}
\usepackage{pxfonts}
\usepackage{graphicx}
\usepackage{color}
\usepackage{float}
\begin{document}

\date{\today}

\title{Long-range spatial extension of exciton states in van der Waals heterostructure}

\author{Zhiwen~Zhou}
\affiliation{Department of Physics, University of California San Diego, La Jolla, CA 92093, USA}
\author{E.~A.~Szwed}
\affiliation{Department of Physics, University of California San Diego, La Jolla, CA 92093, USA}
\author{W.~J.~Brunner}
\affiliation{Department of Physics, University of California San Diego, La Jolla, CA 92093, USA}
\author{H.~Henstridge}
\affiliation{Department of Physics, University of California San Diego, La Jolla, CA 92093, USA}
\author{L.~H.~Fowler-Gerace}
\affiliation{Department of Physics, University of California San Diego, La Jolla, CA 92093, USA}
\author{L.~V.~Butov} 
\affiliation{Department of Physics, University of California San Diego, La Jolla, CA 92093, USA}

\begin{abstract}
\noindent 
Narrow lines in photoluminescence (PL) spectra of excitons are characteristic of low-dimensional semiconductors. These lines correspond to the emission of exciton states in local minima of a potential energy landscape formed by fluctuations of the local exciton environment in the heterostructure. The spatial extension of such states was in the nanometer range. In this work, we present studies of narrow lines in PL spectra of spatially indirect excitons (IXs) in a MoSe$_2$/WSe$_2$ van der Waals heterostructure. The narrow lines vanish with increasing IX density. The disappearance of narrow lines correlates with the onset of IX transport, indicating that the narrow lines correspond to localized exciton states. The narrow lines extend over distances reaching several micrometers and over areas reaching ca.~ten percent of the sample area. This macroscopic spatial extension of the exciton states, corresponding to the narrow lines, indicates a deviation of the exciton energy landscape from random potential and shows that the excitons are confined in moir{\'e} potential with a weak disorder.
\end{abstract}
\maketitle

Narrow lines with linewidths $\lesssim 1$~meV in PL spectra of excitons are ubiquitous in low-dimensional semiconductors. The narrow lines found in a GaAs quantum dot correspond to the emission of exciton states in the dot~\cite{Brunner1992}. The narrow lines observed in GaAs quantum wells originate from the emission of exciton states in local minima of random potential formed by fluctuations of the local exciton environment in the heterostructure, e.g. quantum well width and materials fluctuations~\cite{Zrenner1994, Hess1994, Gammon1996, High2009}. 

Recent studies revealed narrow lines in PL spectra of excitons in van der Waals (vdW) heterostructures composed of single atomic layers of transition-metal dichalcogenides (TMD). The narrow lines observed in a TMD electrostatically defined trap correspond to the emission of exciton states in the trap~\cite{Shanks2022}. The narrow lines were also observed for excitons in local minima of random potential in monolayer TMD~\cite{Srivastava2015, He2015, Koperski2015, Chakraborty2015, Tonndorf2015} and for excitons confined by strain in the regions of the heterostructure flake edges~\cite{Kumar2015}, heterostructure wrinkles~\cite{Branny2016}, or nanopillars~\cite{Branny2017, Palacios-Berraquero2017, Kremser2020}.

Local minima in exciton potential landscape can be also formed in moir{\'e} superlattices in TMD heterostructures~\cite{Wu2018, Yu2018, Wu2017, Yu2017, Zhang2017a}. The moir{\'e} potentials can be affected by atomic reconstruction~\cite{Weston2020, Rosenberger2020, Zhao2023}. Narrow lines were observed for excitons in bilayer TMD heterostructures with moir{\'e} potentials~\cite{Seyler2019, Weijie2020, Bai2020, Baek2020, Baek2021, Liu2021, Brotons-Gisbert2021, Wang2021, Kim2023}. 

Similar to other low-dimensional semiconductors, such as GaAs heterostructures outlined above, narrow lines are ubiquitous in vdW heterostructures. In addition to TMD heterostructures with the exciton confinement caused by electrostatic traps~\cite{Shanks2022}, random potentials~\cite{Srivastava2015, He2015, Koperski2015, Chakraborty2015, Tonndorf2015}, or strain~\cite{Kumar2015, Branny2016, Branny2017, Palacios-Berraquero2017, Kremser2020} and TMD bilayer heterostructures with moir{\'e} potentials~\cite{Seyler2019, Weijie2020, Bai2020, Baek2020, Baek2021, Liu2021, Brotons-Gisbert2021, Wang2021, Kim2023} outlined above, narrow lines were also observed in TMD bilayer heterostructures where moir{\'e} potentials are suppressed by hBN spacers~\cite{Mahdikhanysarvejahany2022} and in TMD trilayer heterostructures~\cite{Bai2023}.

In contrast to random potentials, which are characteristic of semiconductor heterostructures with 2D layers formed by several monolayers (like GaAs heterostructures), or single monolayers (like TMD heterostructures), moir{\'e} potentials are periodic in the heterostructure plane. Their lateral period is typically in the range of ca.~10~nm, exceeding the exciton Bohr radius ca.~1~nm and providing a confining potential for excitons~\cite{Wu2018, Yu2018, Wu2017, Yu2017, Zhang2017a}. Narrow lines are observed both in heterostructures with moir{\'e} potentials and in heterostructures without moir{\'e} potentials, as outlined above, and both these types of heterostructure have disorder potentials. Since both disorder and moir{\'e} potentials can produce the narrow-line emission, the roles of the disorder and moir{\'e} potentials in the origin of the narrow lines remain unclear as outlined, in particular, in recent studies of TMD heterostructures~\cite{Mahdikhanysarvejahany2022}. 

The major difference between moir{\'e} potentials and disorder potentials in the origin of the narrow lines is a spatial ordering for the former. The lateral extension of the narrow lines is given by the lateral extension of the corresponding exciton states. The extension of localized exciton states in a random potential is typically on the order of nanometers, and even extension of delocalized exciton states, given by the mean free path, is typically in the nanometer range for excitons in semiconductor heterostructures as overviewed in Ref.~\cite{Zhou2025} (an observation of anomalously long mean free path ca.~10~$\mu$m was presented in Ref.~\cite{Zhou2025}). In contrast, exciton states in a periodic lattice potential, such as a moir{\'e} potential, can extend over long distances limited by imperfections of the periodic potential in the heterostructure. 

In earlier studies of narrow lines, outlined above, the spatial extension of the narrow lines was limited by the spatial resolution of the optical experiments, ca.~1~$\mu$m. This short extension of exciton states associated with the narrow lines is consistent with random potentials, which include disordered moir{\'e} potentials, in the heterostructures. In this work, we studied narrow lines in PL spectra of spatially indirect excitons (IXs) in a MoSe$_2$/WSe$_2$ vdW heterostructure. We observed that the narrow lines extend over distances reaching several microns. This macroscopic spatial extension of the exciton states, corresponding to the narrow lines, indicates a deviation of the excitonic energy landscape from random potential. An ordering in the local environment of excitons, such as a moir{\'e} potential disordered weakly, is consistent with the observed long-range spatial extension of the exciton states.

\vskip 5mm
{\bf Results and discussions}

In the studied MoSe$_2$/WSe$_2$ vdW heterostructure, the adjacent MoSe$_2$ monolayer and WSe$_2$ monolayer form the separated electron and hole layers and IXs are formed by electrons and holes confined in these separated layers~\cite{Rivera2015}. Twisting between the MoSe$_2$ and WSe$_2$ monolayers with the twist angle $\delta \theta \sim 1.1^\circ$ produces a moir{\'e} potential with the moir{\'e} superlattice period $b \sim a/\delta \theta \sim 17$~nm ($a$ is the lattice constant)~\cite{Wu2018, Yu2018, Wu2017, Yu2017, Zhang2017a}. The HS details and the optical measurements are outlined in Supplementary Information (SI).

\begin{figure}
\begin{center}
\includegraphics[width=8.5cm]{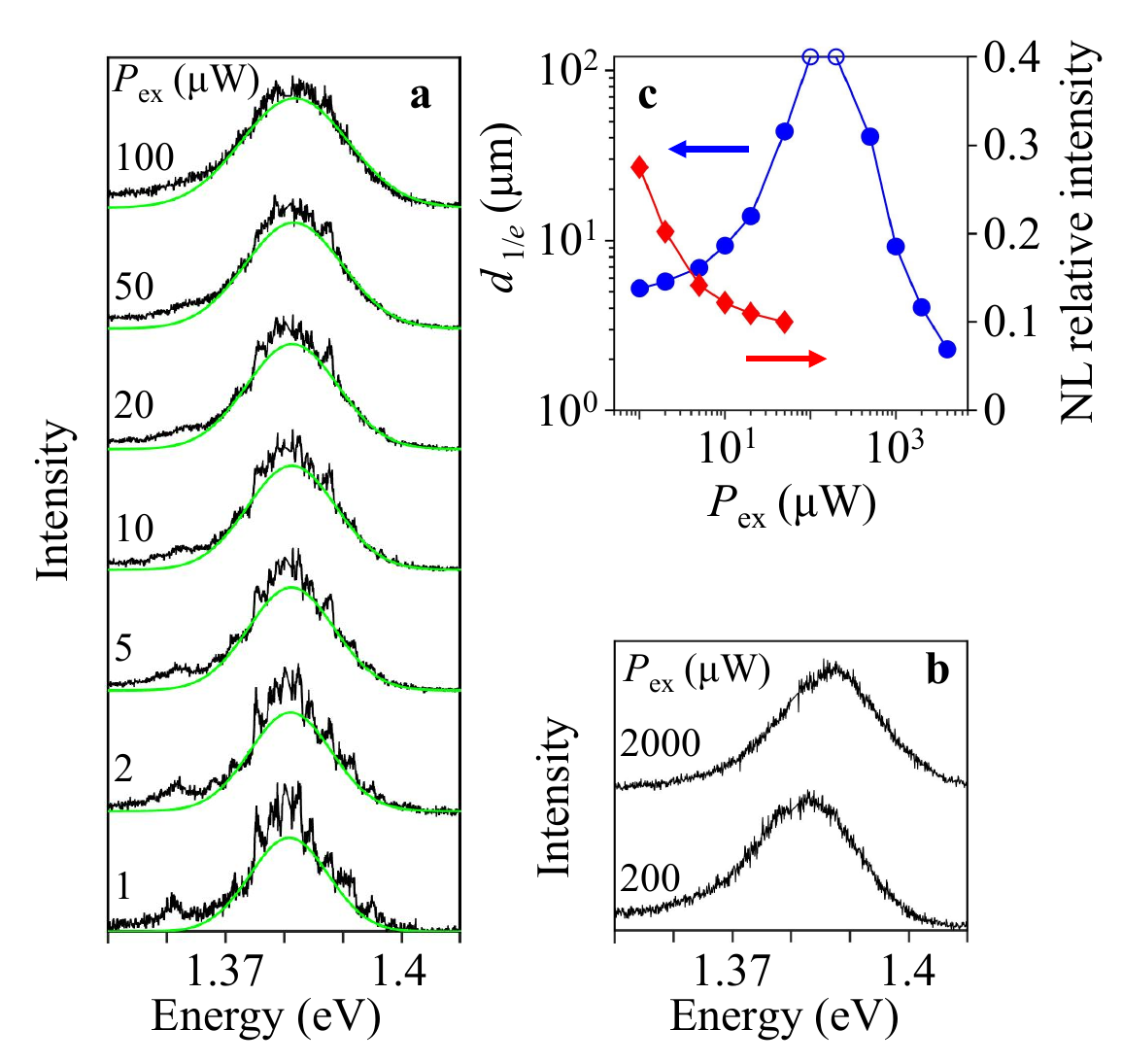}
\caption{Narrow lines in PL spectra of IXs. (a, b) The excitation power $P_{\rm ex}$ dependence of IX spectra. The narrow lines are observed on a background of a broad line approximated by a Gaussian (the green line) for each spectrum. The laser excitation spot is focused to a spot $\sim 2$~$\mu$m in diameter. A higher-energy IX line appearing at $\sim 1.41$~eV at high IX densities was attributed to the appearance of moir{\'e} cells with double occupancy in Ref.~\cite{Zhou2024}. 
(c) Comparison of the relative intensity of the narrow lines (NLs) in the PL spectra with IX transport in the heterostructure. The former is presented by the ratio of the sum of spectrally integrated intensities of the narrow lines to the spectrally integrated intensity of the broad line in the PL spectrum (red diamonds) and the latter is presented by the $1/e$ decay distance of IX transport $d_{1/e}$ from Ref.~\cite{Fowler-Gerace2024} (blue points). 
$T = 3.5$~K.
The disappearance of narrow lines in the spectrum correlates with the onset of IX transport. 
}
\end{center}
\label{fig:spectra}
\end{figure}

{\it Energies of narrow-line exciton states.}
Figure~1 shows narrow lines in PL spectra of IXs in the heterostructure. With increasing density, the energies of the narrow lines stay fixed (Fig.~1), more data on the density dependence of the narrow lines is presented in Fig.~S2 in SI. This indicates that each narrow line corresponds to an exciton state with a low sensitivity to the average exciton density. An exciton in a moir{\'e} cell with certain occupations of neighbor cells is consistent with such a state~\cite{Seyler2019, Weijie2020, Bai2020, Baek2020, Baek2021, Liu2021, Brotons-Gisbert2021, Wang2021, Kim2023}. For such exciton states, adding an exciton to a neighbor cell leads to an increase by the inter-cell interaction energy exceeding the linewidth of the narrow line so, at low densities, cells with statistically distinct occupations of the neighbor cells can produce narrow PL lines with the lack of continuous energy shift with density~\cite{Seyler2019, Weijie2020, Bai2020, Baek2020, Baek2021, Liu2021, Brotons-Gisbert2021, Wang2021, Kim2023}. In contrast, statistical averaging over different exciton states gives the broad PL line with the energy monotonically increasing with average exciton density $n$ (Fig.~1 and Fig.~S2 in SI). This average energy shift can be approximated by the mean field 'capacitor' formula $\delta E = 4 \pi e^2 d_z n/\varepsilon$~\cite{Yoshioka1990} ($d_z \sim 0.6$~nm is the separation between the electron and hole layers and $\varepsilon \sim 7.4$ is the dielectric constant for the heterostructure~\cite{Laturia2018}) and, in particular, can be used for estimating $n$.

\begin{figure*}
\begin{center}
\includegraphics[width=17.5cm]{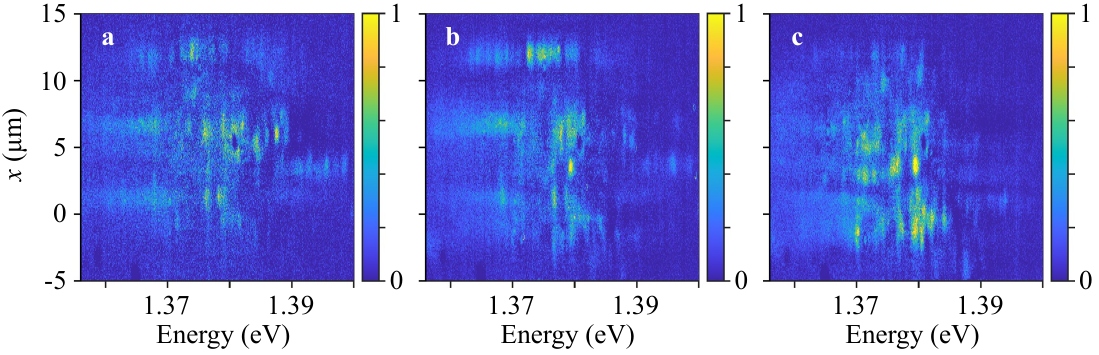}
\caption{$x-$Energy maps of narrow lines. (a-c) $x-$Energy maps of exciton PL for $y = - 3.5~\mu$m (a), $y = - 2.2~\mu$m (b), and $y = - 0.9~\mu$m (c). $x-$Energy maps for other $y$ locations in the heterostructure are shown in Fig.~S4 in SI. The signal is integrated within 1.3~$\mu$m in $y$ direction. The broad background (given by Gaussians in Fig.~1a) is subtracted. 
$x-$Energy maps without background subtraction are shown in Fig.~S5 in SI.
The excitation spot is defocused over a spot $\sim 25$~$\mu$m in diameter covering the heterostructure area for a weak excitation of the entire sample. The excitation power of this defocused excitation is 50~$\mu$W. 
$T = 4.2$~K. 
Narrow lines with longer extensions along $x$ are seen as vertical modulated lines in the $x-$Energy maps.
The narrow lines and, in turn, the corresponding exciton states, seen in the $x-$Energy maps extend over long distances reaching several micrometers. 
}
\end{center}
\label{fig:spectra}
\end{figure*}

{\it Correlation of disappearance of narrow-line exciton states with onset of exciton transport.}
The narrow lines vanish with increasing density and, at high densities, a broad PL line dominates the spectrum (Fig.~1a,b). In Fig.~1c, the relative intensity of the narrow lines in the PL spectra is compared with IX transport in the heterostructure. The former is presented by the ratio of the sum of spectrally integrated intensities of the narrow lines to the spectrally integrated intensity of the broad line in the PL spectrum, and the latter is presented by the $1/e$ decay distance of IX transport $d_{1/e}$ measured in Ref.~\cite{Fowler-Gerace2024}. The opportunity to achieve with varying density both IX localization and long-range IX transport, studied in Ref.~\cite{Fowler-Gerace2024}, enables such comparison. This comparison shows that the disappearance of narrow lines with increasing density correlates with the onset of IX transport (Fig.~1c). The anticorrelation with transport indicates that the narrow lines correspond to localized excitons. However, this anticorrelation does not establish the nature of localization that may be caused by a disorder potential or by an ordered moir{\'e} potential in the heterostructure.

Figure~1 shows that the narrow lines vanish with the onset of IX transport, however, they do not re-appear at the higher densities where IX re-entrant localization, outlined in Ref.~\cite{Fowler-Gerace2024}, is observed. This is consistent with the narrow line association with the exciton localization in local minima of a potential energy landscape formed by variations of the local exciton environment in the heterostructure. The re-entrant localization at the higher densities due to insulating phase, such as the Mott insulator and the Bose glass~\cite{Fisher1989}, is of a different origin. In particular, for the higher densities when most of the moir{\'e} cells are occupied, the particle transport from cell to cell leads to double occupancy that creates a gap for particle-hole excitations, consequently making the state insulating~\cite{Fisher1989}. Figure~1 shows that narrow lines are not characteristic of this high-density insulating phase.

{\it $g$ factor of narrow-line exciton states.}
For all narrow lines, the measured excitonic $g$ factor is $g \sim - 15.5 \pm 0.7$ as shown in Fig.~S3 and outlined in SI. Excitonic $g$ factor is determined by the local atomic registry and the measured $g$ factor corresponds to $H_h^h$ site in the moir{\'e} potential of the MoSe$_2$/WSe$_2$ heterostructure with, in turn, $H$ stacking~\cite{Wu2018, Yu2018, Wu2017, Yu2017, Seyler2019, Wozniak2020}. For TMD heterostructures with moir{\'e} potentials, a coincidence of $g$ factor for all narrow lines was found in Ref.~\cite{Seyler2019}. The $g$ factor specific for a certain local atomic registry ($H_h^h$ in our case) shows that the narrow lines correspond to excitons in the specific site ($H_h^h$ in our case) of the moir{\'e} potential. 

The same local atomic registry $H_h^h$ may extend over a considerable part of the sample in (reconstructed) moir{\'e} potentials that makes the measured $g$ factor essentially insensitive to the location of the exciton state in the sample~\cite{Wozniak2020}. In particular, for the exciton Bohr radius much smaller than the moir{\'e} site, excitons can be localized by random potential fluctuations within the moir{\'e} site as outlined in Ref.~\cite{Mahdikhanysarvejahany2022}. Therefore, the same and atomic-registry-specific $g$ factor of narrow lines is insufficient to establish the nature of localization of the corresponding exciton states that may be caused by a strong disorder or by an ordered moir{\'e} potential in the heterostructure.

\begin{figure*}
\begin{center}
\includegraphics[width=17.5cm]{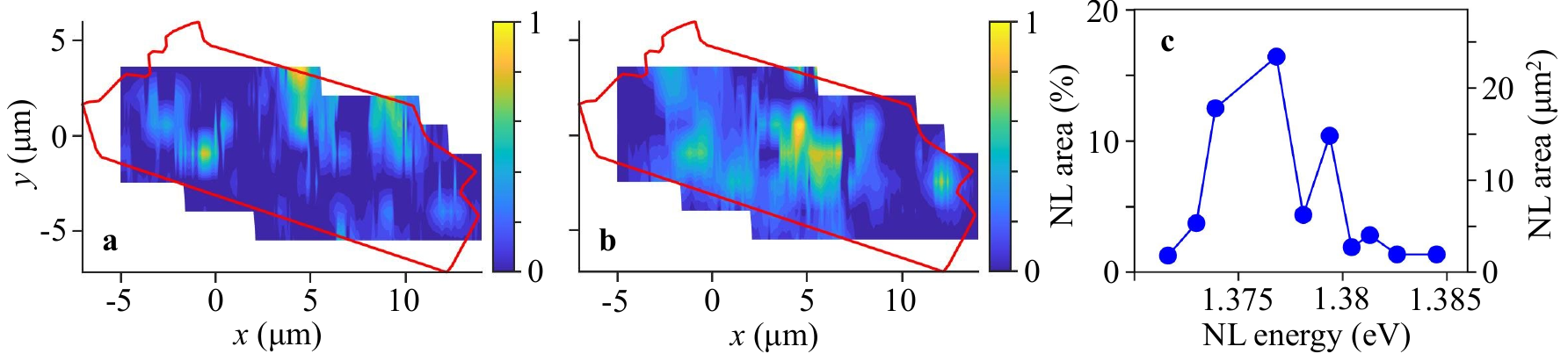}
\caption{$x-y$ maps showing spatial extension of exciton states corresponding to narrow lines. (a, b) $x-y$ maps of spatial extension of the narrow line at $E = 1.3739$~eV (a) and the narrow line at $E = 1.3768$~eV (b). The signal is collected above the broad background (given by Gaussians in Fig.~1a) within the 1~meV linewidth of the narrow line. $x-y$ maps for other narrow lines are shown in Fig.~S6 in SI. The boundary of the MoSe$_2$/WSe$_2$ heterostructure is shown by the red line. The excitation spot is defocused over a spot $\sim 25$~$\mu$m in diameter covering the heterostructure area for a weak excitation of the entire sample. The excitation power of this defocused excitation is 50~$\mu$W. $T = 4.2$~K. 
(c) The area of the exciton state corresponding to the narrow line (NL) vs. the energy of the narrow line. The area boundaries are defined by $1/e$ drop of the narrow-line intensity. The area percentage of the entire measured sample area is also shown.
The narrow lines and, in turn, the corresponding exciton states seen in the $x-y$ maps extend over macroscopic areas reaching ca.~ten percent of the measured sample area. 
}
\end{center}
\label{fig:spectra}
\end{figure*}

{\it Spatial extension of narrow-line exciton states.}
Figure~2 shows $x-$Energy maps of the exciton PL. In these maps, the narrow lines are revealed by the spectrally narrow enhancements of the PL intensity. Figure~2 shows that narrow lines and, in turn, the corresponding exciton states, can extend over long distances reaching several micrometers. 

Figure~2 shows the extension of the narrow-line exciton states in the $x$-direction. The measured $x-$Energy maps at different $y$ locations allow building the $x-y$ maps for the exciton states corresponding to the narrow lines. Examples of the $x-y$ maps for the exciton states are presented in Fig.~3. (Figures~S4 and S6 in SI show $x-$Energy maps for all measured $y$, covering essentially the entire heterostructure area, and $x-y$ maps for many of the narrow-line exciton states seen in the $x-$Energy maps.) Figure~3 shows that the narrow lines extend over distances reaching several micrometers and over areas reaching ca.~ten percent of the measured sample area. 

The observed macroscopic spatial extension of exciton states, corresponding to the narrow lines, indicates a deviation of the exciton energy landscape from random potential. A strong disorder potential does not produce macroscopically extended localized exciton states. In particular, no such extension was observed in any semiconductor system, including GaAs and vdW heterostructures outlined in the introduction, where narrow lines originate from the emission of exciton states localized in local minima of random potential formed by fluctuations of the local exciton environment in the heterostructure, e.g. stress, dielectric, electrostatic, and materials fluctuations. 

In turn, the observed macroscopic spatial extension of localized exciton states, corresponding to the narrow lines, indicates ordering in the local environment of excitons: The exciton state at a certain energy, corresponding to the narrow line, extends over macroscopic length and area (Figs.~2 and 3). A moir{\'e} potential is consistent with such ordering and long-range spatial extension of localized exciton states. We note that different narrow lines and their corresponding exciton states are extended over different regions of the heterostructure (e.g. compare the regions for different narrow lines in Figs.~3a and 3b). This indicates that the local environment for excitons fluctuates over the heterostructure, however, the fluctuations are small enough to allow the long-range extension of the individual exciton states. Therefore, the long-range extension of the narrow lines shows that the excitons are confined in moir{\'e} potential with a weak disorder.

Moir{\'e} potentials with a weak disorder can host long-range ballistic exciton transport due to exciton superfluidity in periodic potentials~\cite{Fisher1989}. A strong disorder destroys superfluidity~\cite{Fisher1989}. Therefore, the weakness of disorder in the moir{\'e} potential, revealed by the long-range extension of the exciton states (Figs.~2 and 3), suggests an opportunity to observe the long-range ballistic transport of excitons in this weakly disordered moir{\'e} potential. Indeed, the studies of exciton transport in the same heterostructure show the long-range ballistic exciton transport over the entire sample with the anomalously long mean free path reaching ca.~10~microns~\cite{Zhou2025}. 

The extension of exciton states over distances reaching several micrometers raises a question of distinguishing such states from delocalized excitons states and a question if such extended states can be called localized states. In this work, we qualitatively discuss an exciton state confined in a region, even of a large area, as a localized state and exciton states, which can travel over different localization regions, as delocalized states. The long-range extension of localized exciton states and their small energy difference facilitates exciton 
transport over different localization regions in the heterostructure.

In summary, we studied narrow lines in PL spectra of IXs in a MoSe$_2$/WSe$_2$ heterostructure. We found that the disappearance of narrow lines correlates with the onset of IX transport, indicating that the narrow lines correspond to localized exciton states. We found that the narrow lines extend over distances reaching several micrometers and over areas reaching ca.~ten percent of the sample area. This macroscopic spatial extension of exciton states, corresponding to the narrow lines, indicates a deviation of the exciton energy landscape from random potential and shows that the excitons are confined in moir{\'e} potential with a weak disorder. The long-range extension of exciton states facilitates efficient exciton transport.

\vskip 5mm
{\bf Acknowledgements}
We thank M.M.~Fogler for discussions and A.K.~Geim for teaching us manufacturing TMD HS. The studies were supported by the Department of Energy, Office of Basic Energy Sciences, under award DE-FG02-07ER46449. The HS manufacturing was supported by NSF Grant 1905478.

\end{document}


\date{\today}

\title{Supporting Information for

Long-range spatial extension of exciton states in van der Waals heterostructure
}

\author{Zhiwen~Zhou}
\affiliation{Department of Physics, University of California San Diego, La Jolla, CA 92093, USA}
\author{E.~A.~Szwed}
\affiliation{Department of Physics, University of California San Diego, La Jolla, CA 92093, USA}
\author{W.~J.~Brunner}
\affiliation{Department of Physics, University of California San Diego, La Jolla, CA 92093, USA}
\author{H.~Henstridge}
\affiliation{Department of Physics, University of California San Diego, La Jolla, CA 92093, USA}
\author{L.~H.~Fowler-Gerace}
\affiliation{Department of Physics, University of California San Diego, La Jolla, CA 92093, USA}
\author{L.~V.~Butov} 
\affiliation{Department of Physics, University of California San Diego, La Jolla, CA 92093, USA}

\begin{abstract}
\end{abstract}
\maketitle

\renewcommand*{\thefigure}{S\arabic{figure}}



\subsection{
Heterostructure}

The MoSe$_2$/WSe$_2$ heterostructure (Fig.~S1a) was assembled using the dry-transfer peel technique~\cite{Withers2015}. The manufacturing details are described in Ref.~\cite{Fowler-Gerace2024} where the same heterostructure was used for studies of IX transport. The thickness of the bottom hBN layer is $\sim 40$~nm, the thickness of the top hBN layer is $\sim 30$~nm. The MoSe$_2$ monolayer is on top of the WSe$_2$ monolayer. The long WSe$_2$ and MoSe$_2$ edges (Fig.~S1b) enable a rotational alignment between the WSe$_2$ and MoSe$_2$ monolayers. The twist angle between the monolayers $\delta \theta = 1.1^\circ$ corresponding to the moir{\'e} superlattice period $b = 17$~nm agrees with the angle between MoSe$_2$ and WSe$_2$ edges in the heterostructure (Fig.~S1b). 

Figure~S1b shows the layer pattern of the heterostructure. The hBN layers cover the entire areas of MoSe$_2$ and WSe$_2$ layers. There was a narrow multilayer graphene electrode on the top of the heterostructure around $x = 2$~$\mu$m for $y = 0$ in Fig.~S1b, this electrode was detached. 

So far, the long-range extension of the narrow PL lines and the corresponding exciton states was realized in one sample in this work. Other studies of GaAs and TMD heterostructures show short extension of the narrow lines, below the spatial resolution of the optical measurements, as outlined in the main text. A shorter extension of exciton states likely originates from stronger disorder. This work demonstrates the existence of the long-range extension of exciton states in TMD heterostructures. Studying this phenomenon in other samples is the subject for future works.

\begin{figure*}
\begin{center}
\includegraphics[width=10.5cm]{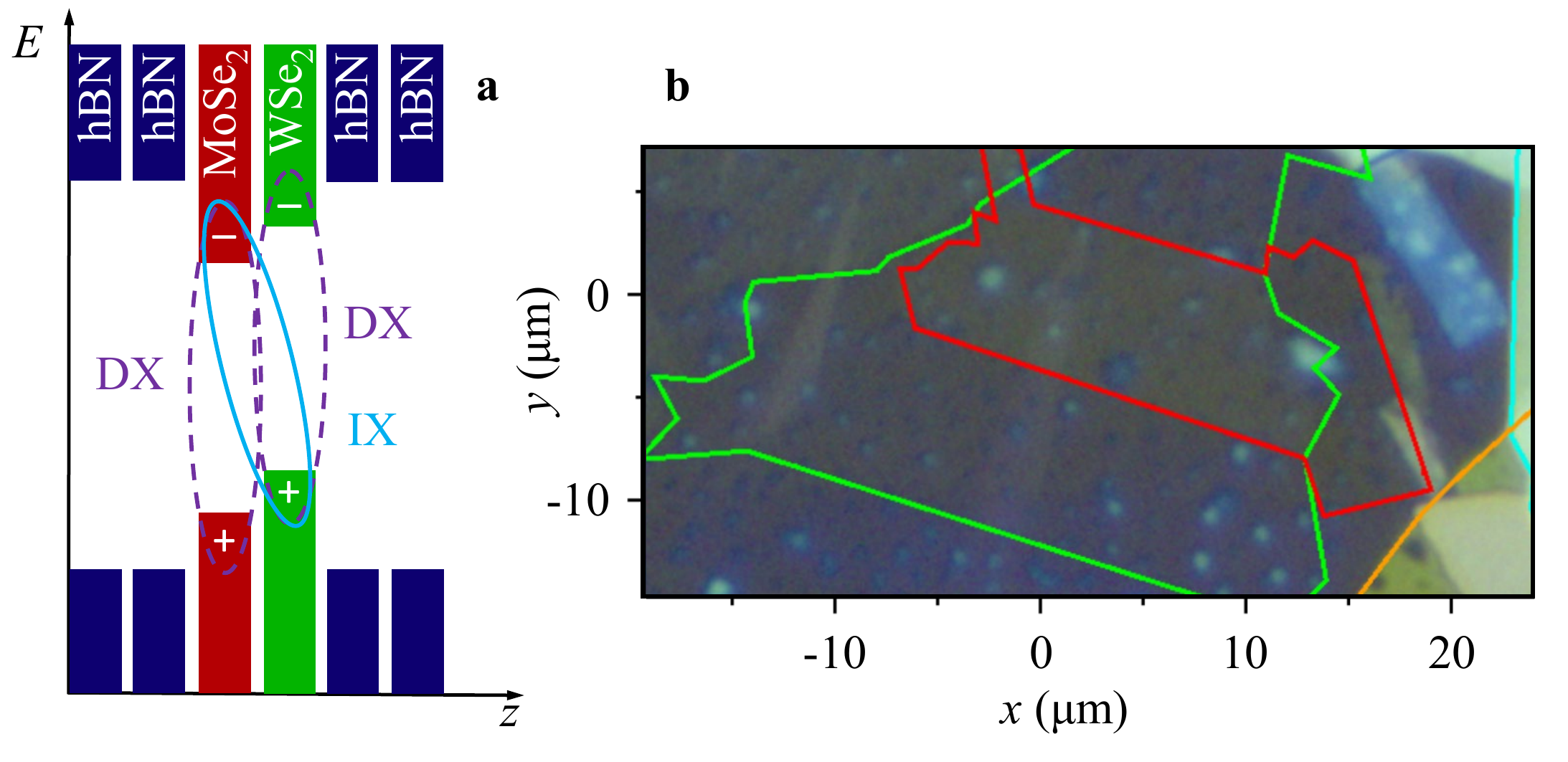}
\caption{(a) Energy-band diagram of the MoSe$_2$/WSe$_2$ heterostructure. The ovals indicate spatially direct excitons (DXs) and spatially indirect exciton (IX) composed of an electron ($-$) and a hole ($+$). (b) A microscope image showing the heterostructure layers. The green, red, cyan, and orange lines indicate the boundaries of WSe$_2$ and MoSe$_2$ monolayers and bottom and top hBN layers, respectively.
}
\end{center}
\label{fig:spectra}
\end{figure*}

\subsection{
Optical measurements}

Excitons were generated by a continuous-wave Ti:sapphire laser with the excitation energy $E_{\rm ex} = 1.689$~eV resonant to DX in WSe$_2$ HS layer. PL spectra were measured using a spectrometer with a resolution of 0.2~meV and a liquid-nitrogen-cooled CCD. The $x-$Energy images were measured with the step 1.3~$\mu$m and the signal integration within 1.3~$\mu$m in the $y$ direction given by a slit (the slit positions in the measurements of the $x-$Energy maps are shown in Fig.~S4h). For the $x-y$ maps, the step in the slit position and the signal integration within the slit 1.3~$\mu$m gave the spatial resolution in the $y$ direction. ${\rm NA} = 0.64$ of the lens gave the spatial resolution 0.7~$\mu$m in the $x$ direction. 

The $g$ factor measurements were performed using circularly polarized excitation and co- and cross-polarized PL signal in magnetic fields up to 8~T oriented perpendicular to the heterostructure plane. The $x-$Energy and $x-y$ images present co-polarized PL. The sample was mounted on an Attocube $xyz$ piezo translation stage allowing adjusting the sample position relative to a focusing lens inside the cryostat.

\subsection{
Density dependence}

Figure~S2 shows the excitation power $P_{\rm ex}$ dependence of IX spectra, similar to the excitation power dependence in Fig.~1a,b in the main text. At low $P_{\rm ex}$, the narrow lines are observed on a background of a broad line. This broad-line background is subtracted from the spectra in Fig.~S2b.

The broad background in this work is approximated by the Gaussians drawn through the origin of the narrow lines as shown in Fig.~1a in the main text. 
In this approach, the spectrum is fitted by a broad Gaussian presenting the broad background (green Gaussian in Fig.~1a) and narrow Gaussians for each narrow line.
In an alternative approach, we drew the broad Gaussians as the fits for the spectra and analyzed the deviations of the spectra from those Gaussians. The differences between the maxima and neighbor minima in the deviations presented the amplitudes of the narrow lines. This alternative approach gave similar results to the approach presented in Fig.~1a.

\begin{figure*}
\begin{center}
\includegraphics[width=12.5cm]{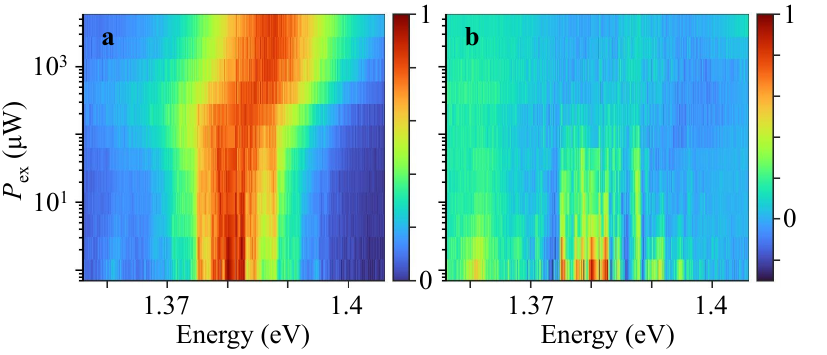}
\caption{(a,b) The excitation power $P_{\rm ex}$ dependence of IX spectra. The spectra intensities at different $P_{\rm ex}$ are normalized. At low $P_{\rm ex}$, the narrow lines are observed on a background of a broad line. This broad-line background is subtracted from the spectra in (b). The laser excitation spot is focused to a spot $\sim 2$~$\mu$m in diameter. $T = 3.5$~K. A higher-energy IX line appearing at $\sim 1.41$~eV at high IX densities was attributed to the appearance of moir{\'e} cells with double occupancy in Ref.~\cite{Zhou2024}. 
}
\end{center}
\label{fig:spectra}
\end{figure*}

\begin{figure*}
\begin{center}
\includegraphics[width=15.5cm]{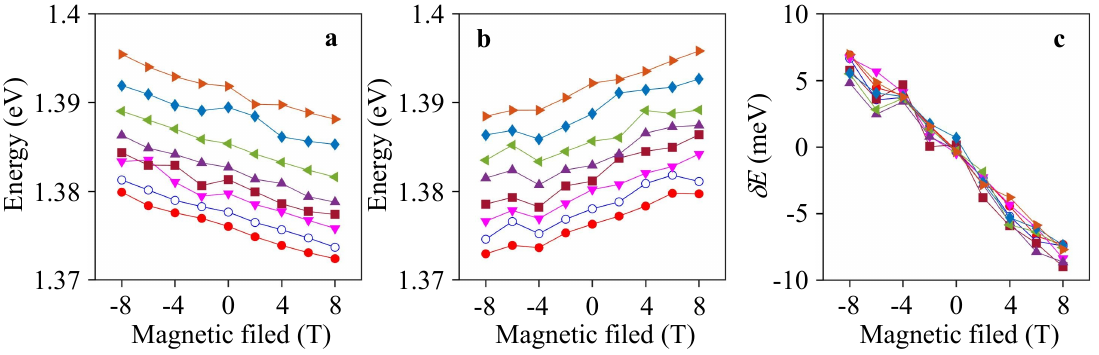}
\caption{$g$ factor of narrow-line exciton states. (a,b) Energies of co-polarized $E_{\rm co}$ (a) and cross-polarized $E_{\rm cross}$ (b) emission of the narrow lines vs. magnetic field. The laser excitation is circularly polarized. (c) The energy difference $\delta E = E_{\rm co} - E_{\rm cross}$ for the narrow lines vs. magnetic field $B$. The same symbol and color is used for a certain narrow line in (a), (b), and (c). 
}
\end{center}
\label{fig:spectra}
\end{figure*}

\begin{figure*}
\begin{center}
\includegraphics[width=17.5cm]{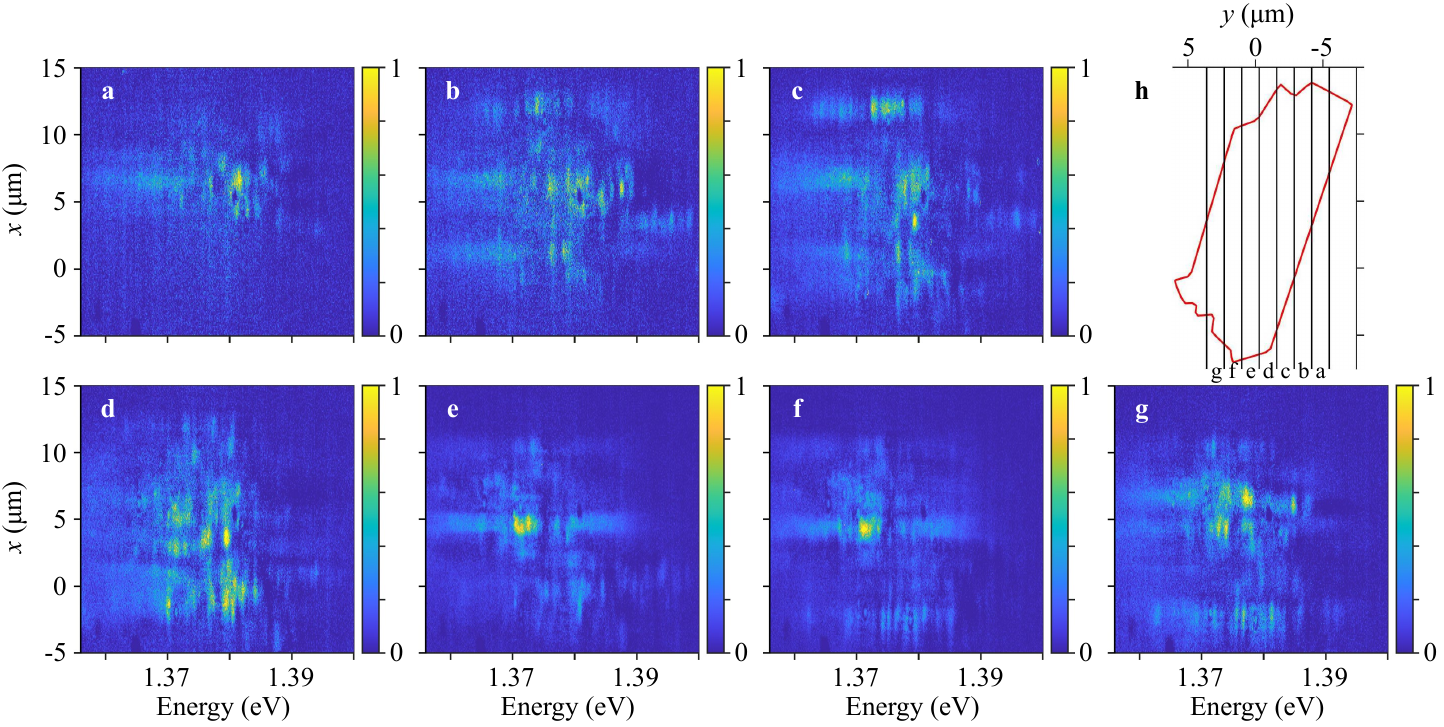}
\caption{(a-g) $x-$Energy maps of narrow lines for $y = - 4.8~\mu$m (a), $y = - 3.5~\mu$m (b), $y = - 2.2~\mu$m (c), $y = - 0.9~\mu$m (d), $y = 0.4~\mu$m (e), $y = 1.7~\mu$m (f), and $y = 3~\mu$m (g). The signal is integrated within 1.3~$\mu$m in $y$ direction. The broad background (given by Gaussians in Fig.~1a in the main text) is subtracted. $x-$Energy maps without background subtraction are shown in Fig.~S5. The excitation spot is defocused over a spot $\sim 25$~$\mu$m in diameter covering the heterostructure area for a weak excitation of the entire sample. The excitation power of this defocused excitation is 50~$\mu$W. $T = 4.2$~K. 
(h) $x-y$ map of the sample showing the $y$ positions of the slit and the 1.3~$\mu$m ranges of the signal integration in the $y$ direction given by the slit for the $x-$Energy maps in (a-g). The boundary of the MoSe$_2$/WSe$_2$ heterostructure is shown by the red line.
}
\end{center}
\label{fig:spectra}
\end{figure*}

\begin{figure*}
\begin{center}
\includegraphics[width=17.5cm]{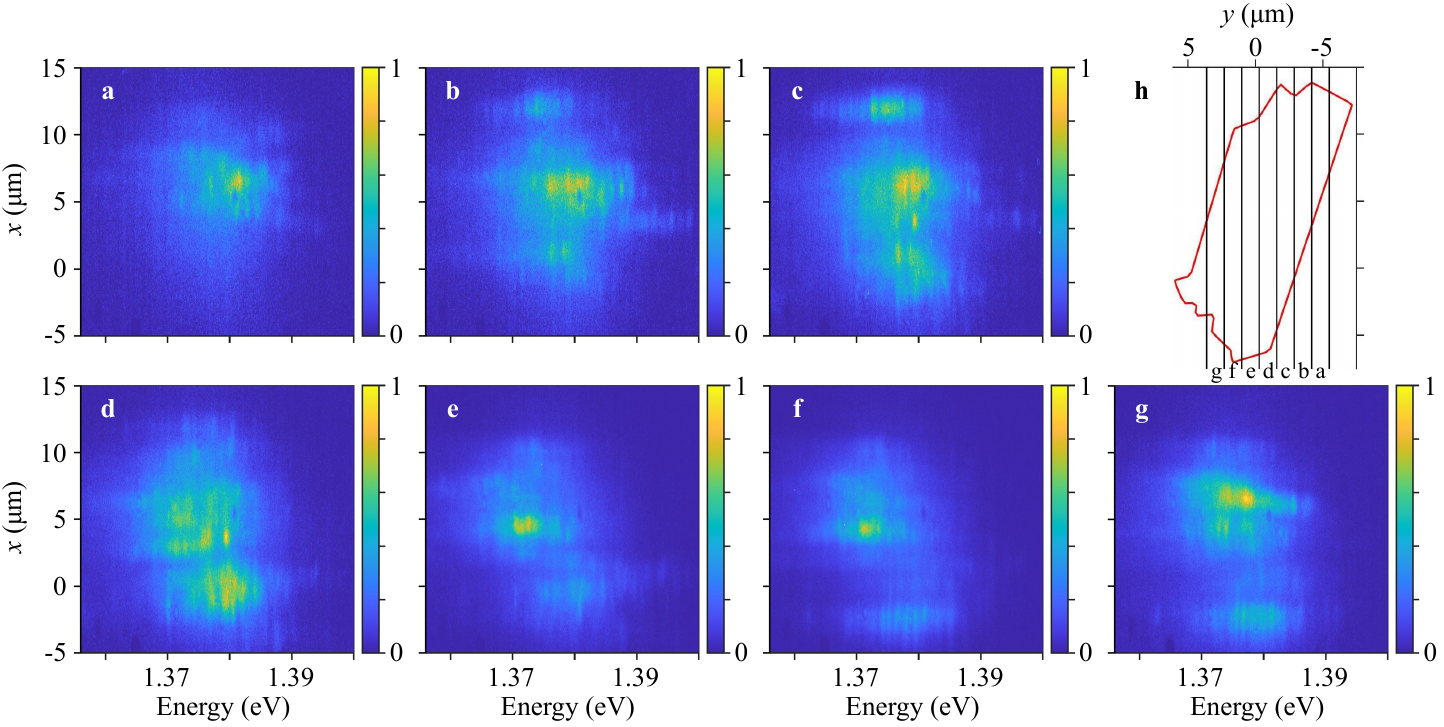}
\caption{$x-$Energy maps of narrow lines similar to the maps in Fig.~S4, however, with no subtraction of the broad background. 
}
\end{center}
\label{fig:spectra}
\end{figure*}

\begin{figure*}
\begin{center}
\includegraphics[width=17.5cm]{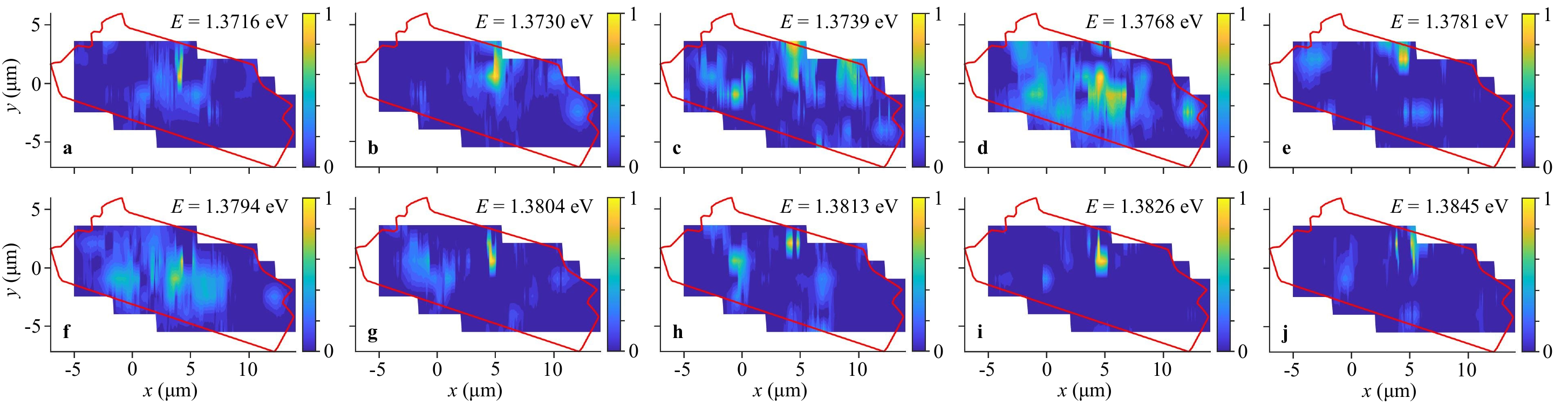}
\caption{$x-y$ maps showing the spatial extension of the exciton states corresponding to the narrow lines. The energies of the narrow lines are indicated. The signal is collected above the broad background (given by Gaussians in Fig.~1a in the main text) within the 1~meV linewidth of the narrow line. The boundary of the MoSe$_2$/WSe$_2$ heterostructure is shown by the red line. The excitation spot is defocused over a spot $\sim 25$~$\mu$m in diameter covering the heterostructure area for a weak excitation of the entire sample. The excitation power of this defocused excitation is 50~$\mu$W. $T = 4.2$~K. 
}
\end{center}
\label{fig:spectra}
\end{figure*}

\begin{figure*}
\begin{center}
\includegraphics[width=17.5cm]{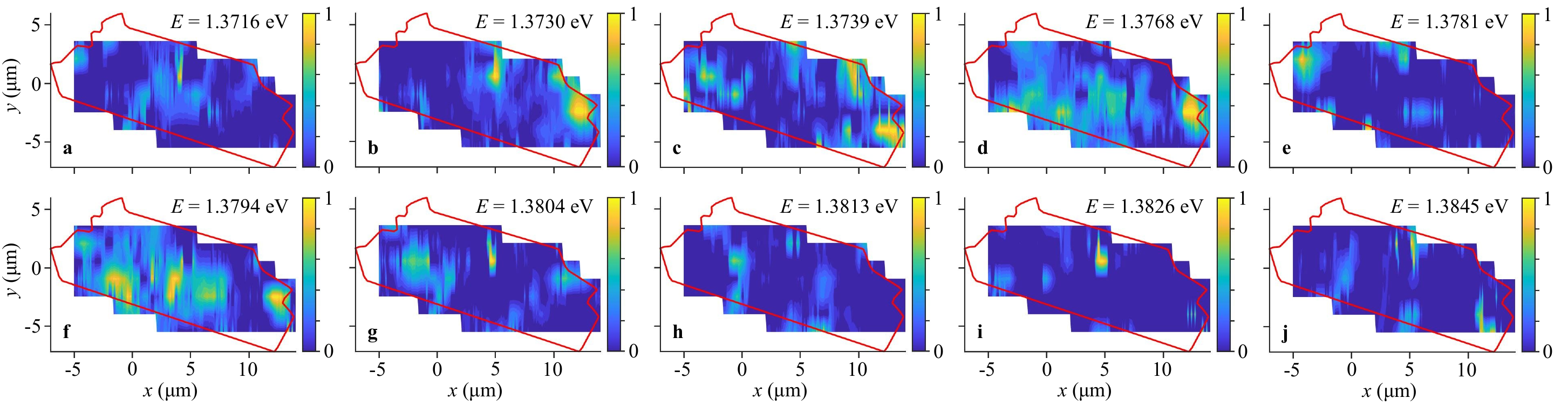}
\caption{$x-y$ maps of narrow lines similar to the maps in Fig.~S6, however, with the intensity of the narrow line normalized by the intensity of the broad-line background at each location.
}
\end{center}
\label{fig:spectra}
\end{figure*}

\subsection{
$g$-factor}

The $g$ factor of the exciton states corresponding to the narrow PL lines is measured using circularly polarized laser excitation and co-polarized and cross-polarized PL of the narrow lines. The co-polarized PL of the narrow lines is $\sim 4$~times stronger than the cross-polarized. Figures~S3a and S3b show the energies of co-polarized $E_{\rm co}$ and cross-polarized $E_{\rm cross}$ PL of the narrow lines vs. magnetic field. Their difference allow estimating the exciton $g$ factor: $\delta E = E_{\rm co} - E_{\rm cross} = g \mu_{\rm B}B$, where $\mu_{\rm B}$ is the Bohr magneton. Figure~S3c shows that for all narrow lines, the measured excitonic $g$ factor is $g \sim - 15.5 \pm 0.7$. 
The measured $g \sim -15.5 \pm 0.7$ corresponds $H_h^h$ site in the moir{\'e} potential of the MoSe$_2$/WSe$_2$ heterostructure with, in turn, H stacking~\cite{Seyler2019, Wozniak2020}.

\subsection{
$x-$Energy maps}

Figure~S4 shows $x-$Energy maps of the exciton PL. This figure is similar to Fig.~2 in the main text, however, it shows $x-$Energy maps for more $y$ locations in the heterostructure. In these maps, the narrow lines are revealed by the spectrally narrow enhancements of the PL intensity. In Fig.~2 in the main text and in Fig.~S4, the broad background (given by Gaussians in Fig.~1a in the main text) is subtracted. $x-$Energy maps without background subtraction are shown in Fig.~S5.

\subsection{
$x-y$ maps}

Figure~S6 shows the $x - y$ maps for the exciton states corresponding to the narrow lines. This figure is similar to Fig.~3 in the main text, however, it shows $x - y$ maps for more narrow lines in the heterostructure.

Figure~S7 shows $x-y$ maps of narrow lines similar to the maps in Fig.~S6. 
For the data in Fig.~S7, the broad background is subtracted as for the data in Fig.~S6. However, while Fig.~S6 shows  the $x-y$ map of intensity of the narrow line, Fig.~S7 shows the $x-y$ map of intensity of the narrow line normalized by the intensity of the broad background at each location. The normalization corrects for signal variations, e.g. due to absorption on the surface, which may contain imperfections.